\documentclass[10pt]{article}

\usepackage{times}
\usepackage{epsfig}
\usepackage{graphicx}
\usepackage{amsmath}
\usepackage{amssymb}
\usepackage{subfig}
\usepackage{float}
\usepackage{multirow,multicol}
\usepackage{breqn}
\usepackage{siunitx}
\usepackage{pdfpages}
\usepackage{enumitem}
\usepackage[margin=1in]{geometry}

\graphicspath{{Figures/}}

\usepackage[pagebackref=true,breaklinks=true,colorlinks,bookmarks=false]{hyperref}

\begin{document}

\title{Unsupervised Medical Image Segmentation with Adversarial Networks: From Edge Diagrams to Segmentation Maps}

\author{Umaseh Sivanesan\\
{\tt\small umaseh.sivanesan@medportal.ca}
\and
Luis H. Braga\\
{\tt\small braga@mcmaster.ca}
\and
Ranil R. Sonnadara\\
Vector Institute, Toronto\\
{\tt\small ranil@mcmaster.ca }
\and
Kiret Dhindsa*\\
Vector Institute, Toronto\\
{\tt\small dhindsj@mcmaster.ca }\\
*\small Corresponding Author\\
Department of Surgery, McMaster University
}

\maketitle

\begin{abstract}
We develop and approach to unsupervised semantic medical image segmentation that extends previous work with generative adversarial networks. We use existing edge detection methods to construct simple edge diagrams, train a generative model to convert them into synthetic medical images, and construct a dataset of synthetic images with known segmentations using variations on extracted edge diagrams. This synthetic dataset is then used to train a supervised image segmentation model. We test our approach on a clinical dataset of kidney ultrasound images and the benchmark ISIC 2018 skin lesion dataset. We show that our unsupervised approach is more accurate than previous unsupervised methods, and performs reasonably compared to supervised image segmentation models. All code and trained models are available at \url{https://github.com/kiretd/Unsupervised-MIseg}.


\end{abstract}

\section{Introduction}

In vivo medical imaging is one of the primary technologies available for clinical evaluation, diagnosis, and treatment planning. The physical challenge of imaging internal tissues is reflected in the low resolution, low signal-to-noise ratio, and high degree of occlusion seen with many common medical imaging technologies. Using medical images to make accurate and meaningful clinical decisions requires substantial training and experience combined with a large body of medical knowledge. As a result, current medical practice places a significant burden on highly trained clinicians specialized in interpreting medical images \cite{Ravi2019,Tajbakhsh2019}. 

A fundamental step in medical image analysis is to identify a region of interest, i.e.,, segmentation. This typically means identifying a bounding region that separates an organ or abnormality from other tissue in the image. For human readers, segmentation allows the extraction of clinically important metrics, such as volume, and for the planning of radiation therapy or surgical removal. In computer-aided-diagnosis (CAD), organ and tissue segmentation allows computer vision models to focus their feature extraction or feature learning computation on the clinically relevant tissue, allowing for more computationally efficient models that are better able to avoid extraneous information in the data \cite{Guo2018}. Manually performing these segmentations is time-consuming, expensive, and subjective, leading to major research effort in developing algorithms that can efficiently perform accurate and reliable semantic segmentation in medical images (i.e.,, segmentation by associating pixels or regions of the image with a classification label).

We present an approach to organ and tissue segmentation based on the use of a Generative Adversarial Networks (GANs) to generate a labelled synthetic training set in the absence of ground truth labels for real medical images. We circumvent the need for labelled real images by generating medical images from simplistic and arbitrary edge diagrams. We then use the synthetic training set to train supervised segmentation models, which are then applied to real images. We evaluate our approach using two datasets: a dataset of ultrasound images for which the task is to segment the kidney, and the ISIC 2018 Skin Lesion Analysis competition dataset, for which the task is to segment skin lesions in dermoscopic images. 

The main contributions of this work are as follows:
\begin{itemize}[noitemsep,nolistsep]
	\item{We demonstrate a novel form of data augmentation by using GANs to generate labelled training data from edge diagrams for applications in which we can exploit a common geometry that is inherent to the semantic segmentation task itself.}
	\item{We show that GANs can generate reasonable synthetic medical images with corresponding organ segmentation maps from just edge diagrams.}
	\item{By generating data using edge diagrams, we show that We can obtain accurate and reliable organ segmentation in a fully unsupervised way, with the option of semi-supervised training if labelled data are available.}
\end{itemize}

\section{Related Work}

Traditional approaches to algorithmic image segmentation were largely unsupervised, i.e.,, they did not rely on ground truth (clinician-supplied) segmentations to train a model. A variety of such methods were developed in previous decades (for example, methods based on edge detection \cite{Canny1987}, region growing \cite{Adams1994}, contour modelling \cite{Kass1988}, and texture analysis \cite{Manjunath1991}); however, these typically relied on built-in constraints about object appearance or differences in contrast or intensity between regions of interest and background pixels. Such constraints do not always work well for medical images, particularly for imaging modalities that produce lower quality images (e.g.,, ultrasound imaging), or for regions of the body where multiple organ and tissue types are imaged together. 

To overcome the shortcomings of these earlier approaches, modern image segmentation techniques often rely on supervised learning with deep neural networks and large amounts of labelled training data. These models are capable of performing semantic segmentation, and thus can categorize regions of images based on meaningful labels provided by a clinician. The most common approach of the last few years has been based on convolutional neural networks (CNNs), which have been widely demonstrated to be successful for many kinds of computer vision tasks \cite{Voulodimos2018, Yoo2015, Garcia2017, Lu2017}. 

Key developments in semantic segmentation have been based on variations of CNNs. The Fully Convolutional Network (FCN) omitted the fully connected layers used in standard CNNs, which are used to obtain a pixel-wise grouping label, and instead used deconvolution layers to obtain segmentation probability maps for images \cite{Long2015}. A similar idea based on encoder-decoder networks was developed by deconvolving VGG16 \cite{Simonyan2014}, a CNN pretrained on the ImageNet dataset \cite{Deng2009} that is sometimes used as a starting point for specific medical imaging problems (e.g., \cite{Lopez2017, Kieffer2017}). In order to take greater advantage of the spatial correlations between pixels that should be grouped together, CNNs have also been combined with Conditional Random Fields (CRFs) \cite{Lin2016}. 

Different forms of these CNN approaches have dominated the field of medical imaging segmentation as well \cite{Moeskops2016, Zhang2015, Kleesiek2016, Dou2016, Brosch2016, Cirecsan2013}. In particular, a specific instance of FCNs, U-net \cite{Ronne2015}, has performed well for a variety of medical imaging segmentation tasks (e.g.,, \cite{Christ2016}). Due to its success, it has since been extended in many ways: for 3D images \cite{Cciccek2016}, with an attention mechanism \cite{Oktay2018}, with a pretrained VGG11 encoder \cite{Iglovikov2018}, and so on. 

Two major limitations reduce the utility of the CNN approaches described above: 1) they are trained explicitly to minimize pixel-wise segmentation error and therefore typically require significant post-processing of their outputs in order to obtain solutions that are spatially contiguous, and 2) they require ground truth segmentations for training, which can be very difficult and expensive to obtain on the scale that is required for effective deep learning. While some recent methods have been able to address the first limitation by training for scalable spatial coherence using patch learning with multi-scale loss functions \cite{Havaei2017, Kamnitsas2017, Pereira2016}, they do not address the need for manually segmented training images. Here we propose the use of a GAN to create synthetic training data that can be used to train supervised image segmentation models when no labelled training data are available, thus allowing for unsupervised medical image segmentation. 

GANs have been formulated as image-to-image translation architectures that take paired images as input \cite{Isola2017}, and thus have been successfully applied to semantic segmentation by training them on pairs of images with their corresponding ground truth segmentations. This has been done in a fully supervised manner \cite{Luc2016,Xue2018,Son2017,Yang2017,Moeskops2016} and in a semi-supervised or weakly supervised manner \cite{Hong2015}. Most interestingly, researchers have taken advantage of the fact that GANs, by their nature, can be used to generate synthetic data as a form of data augmentation \cite{Shin2018}. Using this approach, GANs can be trained with a relatively low number image-segmentation pairs to generate additional training data for a DualGAN semantic segmentation model \cite{Souly2017,Guibas2018}, or a fully supervised model like U-net \cite{Shin2018}. However, these approaches, like the previous CNN-based models, are limited by the fact that ground truth segmentations are required to train the GANs in the first place. 

Different approaches have been taken to overcome the need for segmentation labels during training. W-net pairs two U-nets to form a deep auto-encoder that can be used in combination with a CRF algorithm for scene decomposition \cite{Xia2017}. In contrast, co-segmentation approaches exploit feature similarity for multiple instances of same-class objects in an image, which is suitable for certain kinds of segmentation tasks with distinct ROIs \cite{Hsu2019}. Recent recomposition approaches based on generative modelling (e.g., SEIGAN \cite{Ostyakov2019}) segment foreground objects by moving them to similar background images. Perhaps most similar to ours is a very recent approach, ReDO \cite{Chen2019}, that performs scene decomposition following region-wise composition using a GAN based on the assumption that different objects composing a scene would be statisticaly independent with respect to certain properties, such as colour and texture.

All of the above approaches assume that the target ROI for segmentation is easily distinguishable from the rest of the image along some feature dimensions, such as brightness or colour, and therefore try to define or learn the properties that distinguish regions of the image. In many medical imaging applications, this is extremely difficult to do, as there may not be a set of learnable properties that support the task. In the case of organ segmentation, as demonstrated with the kidney ultrasound dataset presented here, a clinical expert would typically rely heavily on prior anatomical knowledge and experience, which provides an expectation of the contours of the kidney in the absence of a clear boundary. For this reason, non-expert humans are likely to fail at this particular task (see Figure \ref{fig:kidney_grades}). We overcome this challenge using a generative process to learn an expectation of the shape of the ROI in the data generation phase, as described below.

\section{Methods}

\subsection{Overview of our Approach}
Here we propose a way of extending this previous work to generate synthetic training data using GANs in a fully unsupervised way for applications in which there is an expected segmentation geometry that can serve as a prior. It is based on the assumption that there exists a simple template structure that can be exploited to generate simple diagrams with known segmentations, what we call edge diagrams, from which a GAN can generate sufficiently realistic (and similarly challenging) training images. As long as reasonable edge diagrams can be extracted from the original images to train the GAN, and new edge diagrams can be constructed using variations on the template structure as the ground truth segmentations, then synthetic training data can be generated with known segmentations.

Our approach follows a simple recipe. First we generate simple edge diagrams from real unlabelled training images using available computer vision techniques. We use the corresponding image-diagram pairs to train a GAN to produce synthetic medical images from the edge diagrams. We then use a simple algorithm to generate variations of these edge diagrams with known ROIs, and use the trained GAN to synthesize new images from these new edge diagrams. Finally, we use these new purely synthetic image-segmentation pairs to train a supervised image segmentation model that can be used to identify ROIs in real medical images. The entire approach is illustrated in Figure \ref{fig:pipeline}.

\subsection{Dataset 1: Renal Ultrasound Images}
\begin{figure}[t]
\begin{center}
\setlength{\tabcolsep}{1.5pt}
	\begin{tabular}{cc}
    		\subfloat[Grade 1]{\includegraphics[width=0.22\textwidth]{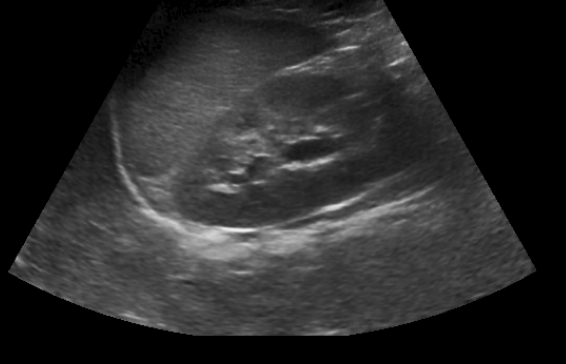}} & 
    		\subfloat[Grade 2]{\includegraphics[width=0.22\textwidth]{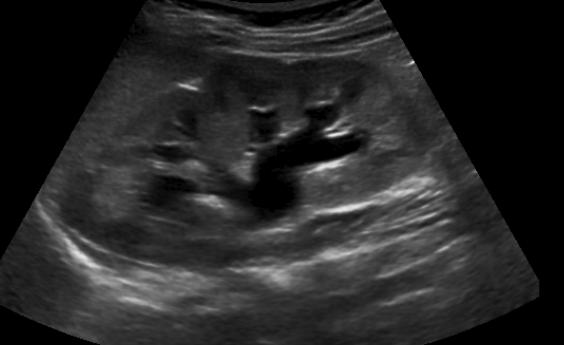}} \\[-2ex]
    		\subfloat[Grade 3]{\includegraphics[width=0.22\textwidth]{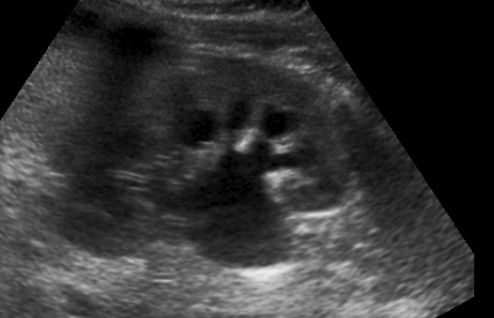}} & 
    		\subfloat[Grade 4]{\includegraphics[width=0.22\textwidth]{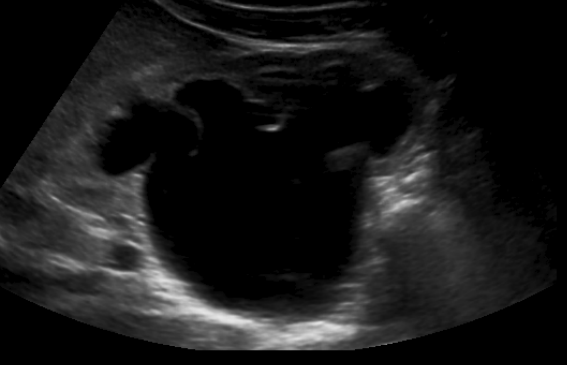}} 
	\end{tabular}
    \caption{Examples from the kidney ultrasound dataset with different hydronephrosis severity grades, from 1 (low severity) to 4 (severe hydronephrosis). }
    \label{fig:kidney_grades}
\end{center}
\end{figure}

We use a dataset of renal ultrasound images developed for prenatal hydronephrosis, a congenital kidney disorder marked by excessive and potentially dangerous fluid retention in the kidneys \cite{Dhindsa2018}. The dataset consists of 2492 2D sagittal kidney ultrasound images from 773 patients across multiple hospital visits. This is a difficult dataset for image segmentation due to poor image quality, unclear contours of the kidneys, and the large variation introduced by different degrees of the kidney disorder called hydronephrosis (see Figure \ref{fig:kidney_grades}). In addition, a major challenge of this dataset is that the two most salient boundaries, the outer ultrasound cone inherent to ultrasound imaging with a probe, and the dark inner region of the kidney, which is caused by fluid retention in hydronephrosis, are both misleading with respect to segmenting the kidney. 


\subsection{Dataset 2: Skin Lesion Segmentation}
We use the ISIC 2018 Challenge dataset to evaluate our model with respect to Task 1 of the challenge: Lesion Boundary Segmentation \cite{Codella2019,Tschandl2018}. By showing that our approach is also successful on this benchmark dataset, we show that the method is not limited to only one domain and imaging modality.

\subsection{Image Preprocessing}
\begin{figure*}
	\centering
	\includegraphics[width=\textwidth]{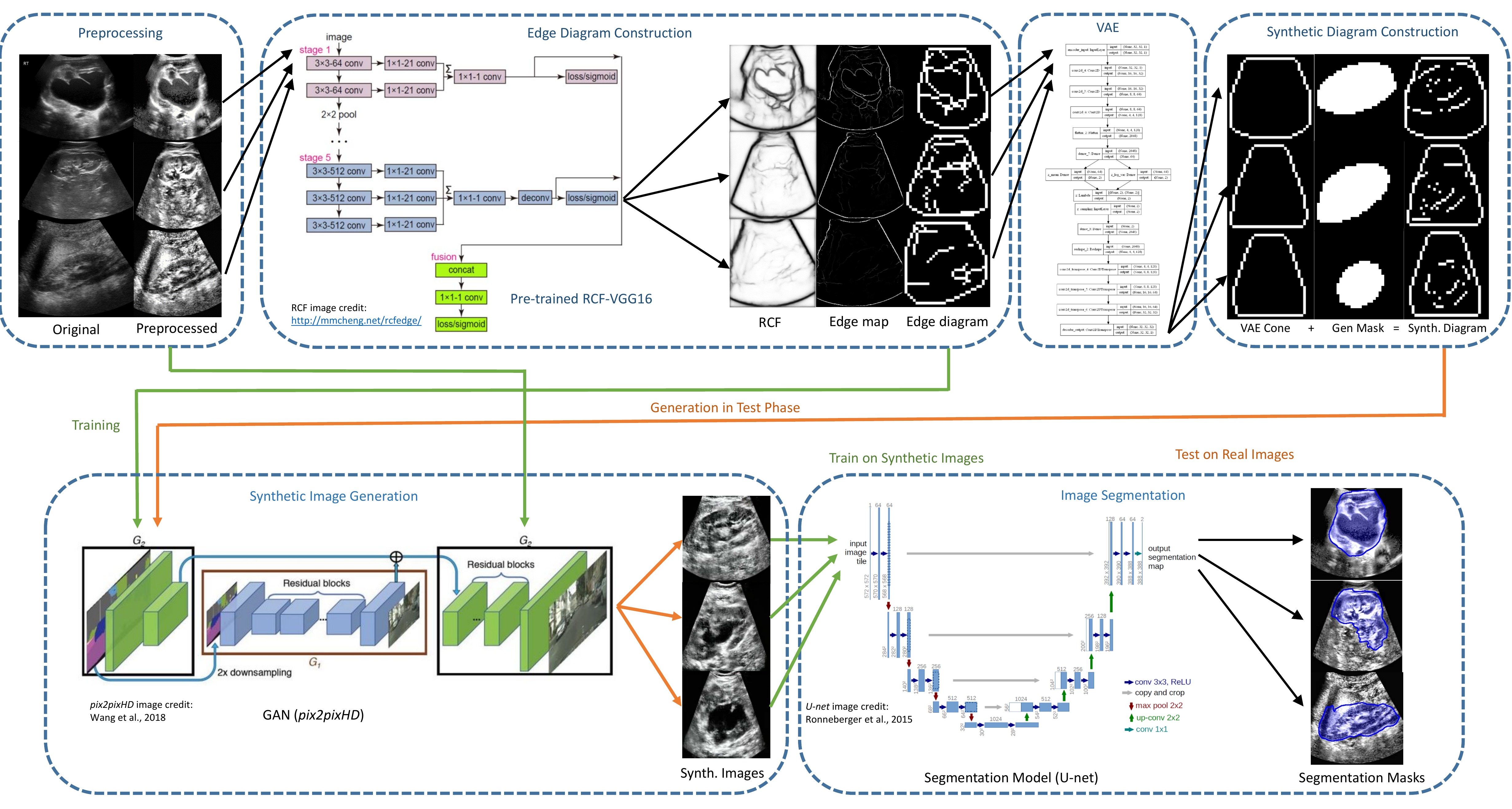}
	\caption{The proposed unsupervised image segmentation pipeline.}
	\label{fig:pipeline}
\end{figure*}

We follow a similar methodology used for preprocessing renal ultrasound imaging for deep learning described in \cite{Dhindsa2018}. We crop the images to remove white borders, despeckle them to remove speckle noise caused by interference with the ultrasound probe during imaging \cite{Tay2010}, and re-scale to 256$\times$256 pixels for consistency. We remove text annotations made by clinicians using the pre-trained Efficient and Accurate Scene Text Detector (EAST) \cite{Zhou2017}. We then normalize the pixel intensity of each image to be from 0 to 1 after trimming the pixel intensity from the 2\textsuperscript{nd} percentile to the 98\textsuperscript{th} percentile of the original pixel intensity across the image. In addition, we enhance the contrast of each image using Contrast Limited Adaptive Histogram Equalization with a clip limit of 0.03 \cite{Pizer1987}. Finally, we normalize the images by the mean and standard deviation of the training set during cross-validation. The results of preprocessing can be seen in the example given in Figure \ref{fig:pipeline}.

We perform no preprocessing for the ISIC skin lesion images other than to resize them to 265 $\times$ 256 pixels. 

\subsection{Creating Edge Diagrams for Training}
\subsubsection{Ultrasound Images}
To obtain edge diagrams from real medical images, we start with a rough edge map given by a pre-trained edge detector \cite{Liu2017} that uses richer convolutional features (RCF) with the VGG16 architecture \cite{Simonyan2014}, which we then fine-tune using non-maximum suppression with Structured Forests for \cite{Dollar2014} edge thinning (as recommended by the authors of RCF). In order to simplify the edge map and remove non-zero pixels that do not belong to the ROI, we downscale the image to 32$\times$32 pixels, remove any regions with an area smaller than 3 pixels, and skeletonize the image \cite{Zhang1984}. 

Since the edge diagrams are simplistic, synthetic edge diagrams can be generated in a variety of ways (e.g.,, they can be drawn by hand if desired). For the results presented here, we train a Variational Autoencoder (VAE) \cite{Kingma2013} to learn a latent space representing edge diagrams obtained from real images. While this model can generate synthetic edge diagrams, it does not directly provide a known ground truth segmentation. We therefore use Otsu's method \cite{Otsu1979, Sezgin2004} for edge detection to extract just the outer profile of the edge diagram, which corresponds to the ultrasound cone (the outer profile of ultrasound images produced by the ultrasound probe). We then generate a ground truth segmentation inside the cone of the synthetic edge diagram to ensure that we know every pixel belonging to the desired segmentation mask. 

To generate the ground truth segmentation representing the kidney ROI, we compute a random ellipse with a random origin, rotation, and major and minor axes within the bounds of the 32$\times$32 pixel edge diagram. We draw randomly selected arcs from the ellipse so as to leave gaps in the kidney outline, simulating occlusion of the kidney boundary. We then also draw an arc inside the ellipse roughly parallel to the major axis to represent the renal pelvis. Finally, we add some noise in the form of random pixels inside the ellipse. Both the extracted and synthetic edge diagrams are rescaled up to 256$\times$256 pixels for training the GAN.

Note that in many medical imaging applications, the entire process involving the VAE may be skipped and only the ground truth segmentation is needed (as we do with the ISIC 2018 dataset). We specifically include the cone and the segmentation in the synthetic edge diagrams for ultrasound images to ensure the GAN generates synthetic ultrasound images with cone profiles, thus preventing the later segmentation model from learning to only segment the outer cone. 

\subsubsection{Other Medical Images}
The same process was used to generate synthetic edge diagrams for the skin lesion images, with two notable exceptions: no cone was created, and random lines, arcs, and smaller ellipses were added to some synthetic edge diagrams to mimic the presence of rulers, pen marks, hairs, and other objects that sometimes appeared in the real images. 

In principle, any method that produces edge diagrams with known segmentations and enough variation can be used. The methods described here are included for reproducibility rather than methodological necessity.

\subsection{Generative Adversarial Networks}
The conventional GAN \cite{Goodfellow2014} uses the loss function

\begin{dmath}
    \min_{\theta_G}\max_{\theta_D}\mathcal{L}(\theta_G,\theta_D) = \mathbb{E}_{x\sim P_X}\left[log(D(x))\right] + \mathbb{E}_{z\sim P_Z}\left[log(1-D(G(z)))\right],
\label{eq:ganloss}
\end{dmath}

where $\theta_G$ and $\theta_D$ are the parameters for generator $G$ and discriminator $D$, $x\in X$ is a real image from our set of real ultrasound images $X$ with unknown distribution $P_X$, and $z$ is a random vector noise vector drawn from some defined probability distribution $P_Z$ (in this case, a Gaussian distribution). Training the GAN involves setting the generator an discriminator in competition with one another: the generator is trained to minimize the objective function by generating images that are indistinguishable from the real training images, and the discriminator is trained to maximize the objective function by learning to distinguish the images synthesized by the generator from real training images. For this work we use the \texttt{pix2pixHD} architecture \cite{Wang2018}.

\subsubsection{\texttt{pix2pixHD} Architecture}
This architecture uses two subnetworks to create a coarse-to-fine generator that can upscale image quality during image-to-image translation, and three multiscale discriminators to address the need to discriminate between high resolution synthetic images and real images while keeping the network size and memory requirements relatively low. Training the entire network comes with a loss function extended from \ref{eq:ganloss} for multiple discriminators by summing over the discriminators to obtain 

\begin{dmath}
    \min_{\theta_G} \left(\left( \max_{\theta_{D_1},\theta_{D_3},\theta_{D_3}} \sum_{k=1}^3 \mathcal{L}(\theta_{G},\theta_{D_k}) \right) + \lambda \sum_{k=1}^3 \mathcal{L}_{FM}(\theta_G,\theta_{D_k})  \right),
\label{eq:pix2pixhd_loss}
\end{dmath}
where $\lambda$ is a parameter used to balance the influence of each term of the loss function. Here, $\mathcal{L}_{FM}$ is the layer-wise feature matching loss that is incorporated to account for the fact that the generator must now model data distributions at multiple scales:

\begin{equation}
    \mathcal{L}_{FM} = \mathbb{E}_{(\mathbf{Z},\mathbf{x})} \sum^T_{i=1}\frac{1}{N_i}\left[\lVert D^{(i)}_k(\mathbf{z},\mathbf{x}) - D^{(i)}_k(\mathbf{z},G(\mathbf{z}) \rVert_1\right],
    \label{eq:featurematching}
\end{equation}
where $T$ is the number of layers and $N_i$ is the number of units in layer $i$. In this work, we are not upscaling the resolution of images, but we find \texttt{pix2pixHD} to also be valuable for translating from a simple image (our edge diagrams) to more complex images (medical images).

\subsection{Training}
\subsubsection{GAN}
For our ultrasound images, a trained surgical urologist provided segmentations for 491 images (approximately evenly split by class; range: 96-100). We reserve those images for evaluation (i.e.,, they are not used to train any model). We additionally remove any training images taken from the same patients that are also represented in the evaluation set to avoid overfitting due to subject-specific characteristics. In total, we use 918 images to train the GAN with 20\% used for validation. From these, we create a synthetic training set of 2000 image-segmentation pairs.

For the ISIC 2018 dataset, 2075 images are used for training and 519 are used for evaluation. Using these data, we generate 3000 synthetic image-segmentation pairs for training and 750 for validation. For both datasets, we generated as many images as required until segmentation accuracy on the validation set no longer improved.

We train our implementation of \texttt{pix2pixHD} using the same settings given in \cite{Wang2018} and choose the parameters corresponding to the epoch that minimizes the Fr\'{e}chet Inception Distance (FID) with respect to the validation data \cite{Heusel2017}. This is the 90th epoch for the ultrasound dataset, and the 100th epoch for the skin lesion dataset.

\subsubsection{VAE}
\begin{figure}
	\centering
	\includegraphics[width=0.4\textwidth]{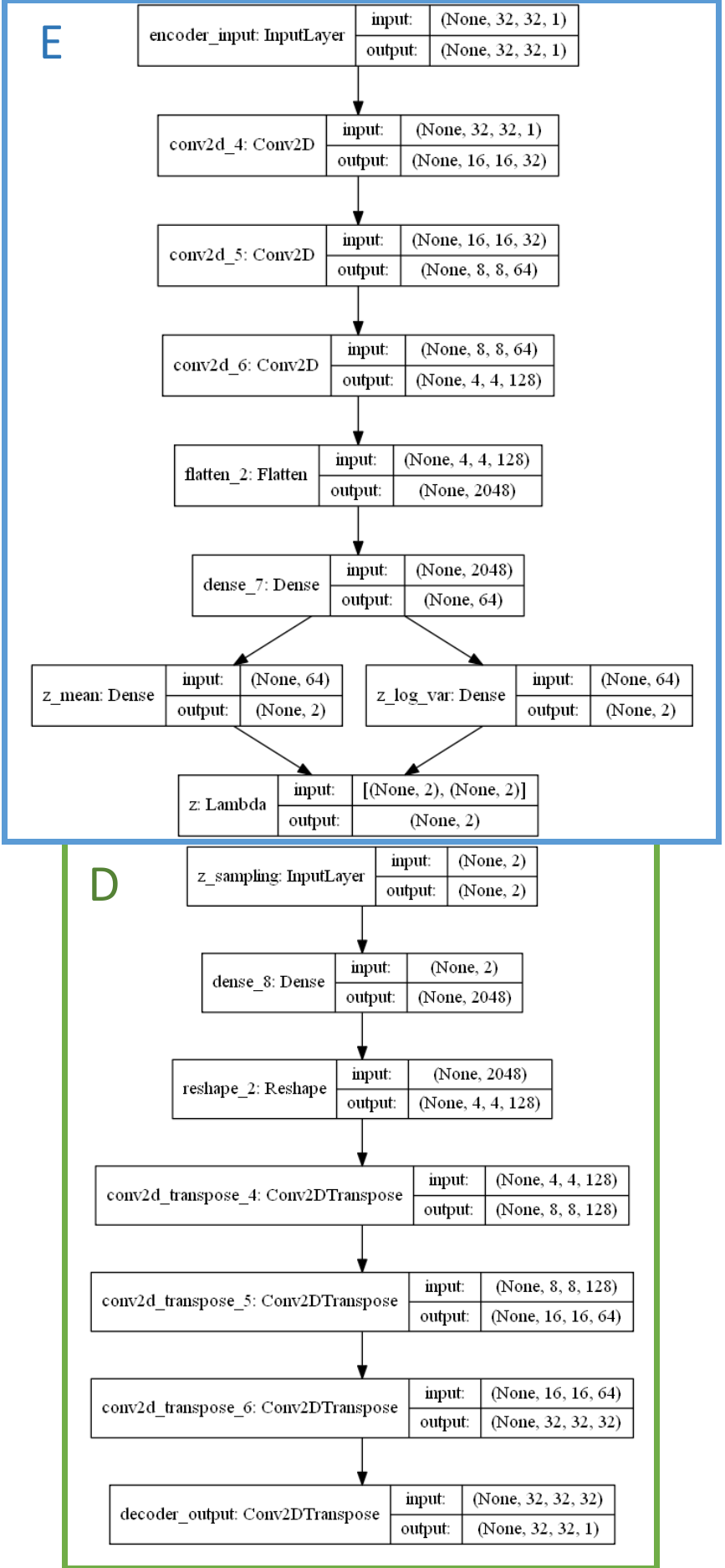}
	\caption{VAE model architecture for generating ultrasound cones.}
	\label{fig:vae}
\end{figure}

The VAE we use to generate ultrasound cones for synthetic edge diagrams is shown in Figure \ref{fig:vae}. We train the VAE over 40 epochs and a batch size of 128 using the adaptive moment estimator (Adam) \cite{Kingma2014} and the Kullback-Leibler divergence loss. 

%
\subsubsection{U-Net}
We use the U-net architecture defined in \cite{Ronne2015} to train a segmentation model for the ultrasound dataset. However, we use the sum of the pixel-wise binary cross-entropy and the dice coefficient as our loss function. We use Adam for optimization with a batch size of 1. Finally, we perform data augmentation with horizontal flips (50\% probability) and horizontal and vertical translations of up to 26 pixels (10\%). 


\subsubsection{Mask-RCNN}
We use the Mask-RCNN implementation provided here \cite{maskrcnn} adjusted for the ISIC 2018 dataset. We use anchor sizes of $2^{i}, i\in\{3, 4, 6, 7, 8\}$ and 32 training ROIs per image. Other hyperameters were kept as default values. We perform data augmentation with both horizontal and vertical flips (50\% probability), rotation of \ang{90} or \ang{270}, and a Gaussian blur of up to 5 standard deviations. 

\subsection{Evaluation}
We evaluate our model using three standard metrics: the F1 score, mean intersection over union (mIoU), and pixel-wise classification accuracy (pACC). 

\subsubsection{Comparison with W-net}
We train W-net with the soft normalized cut term in the loss function \cite{Xia2017}. In addition, we perform the recommended post-processing of the W-net generated segmentation maps using a fully-connected CRF for edge recovery, and hierarchical image segmentation for contour grouping \cite{Arbalaez2011}.

\subsubsection{Mask Extraction from Clinician-Provided Kidney Segmentations}
\begin{figure*}[t!]
	\centering
	\begin{tabular}{c}
    		\subfloat[Successful mask extraction]{\includegraphics[width=0.8\textwidth]{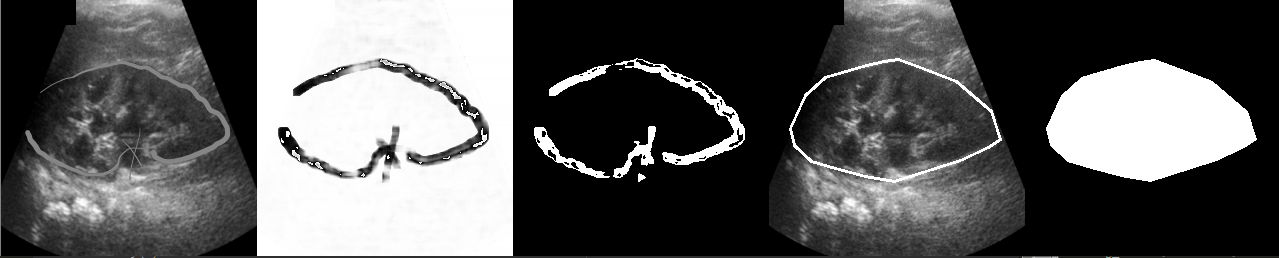}} \\[-1em]
    		\subfloat[Unsuccessful mask extraction]{\includegraphics[width=0.8\textwidth]{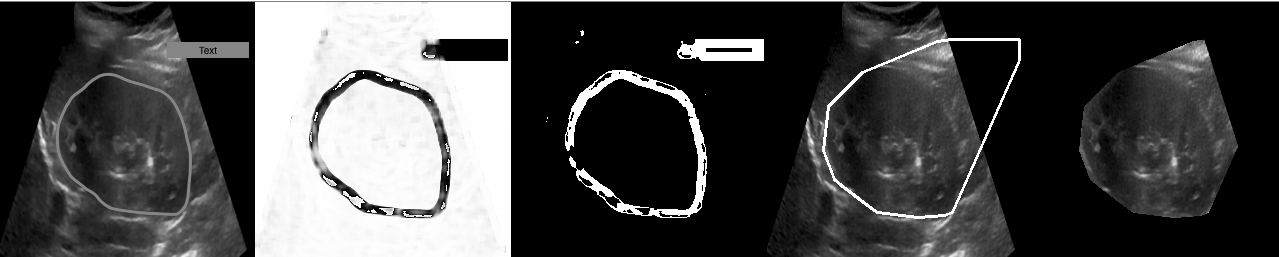}}
	\end{tabular}
	\caption{A successful and unsuccessful example of mask extraction from clinician-provided kidney segmentations. From left to right, panel 1 shows the original image with the kidney outlined by the clinician, panel 2 shows the difference between panel 1 and the original image from our database without the outline, panel 3 shows the difference image thresholded by pixel value, panel 4 shows the convex hull of the thresholded image in panel 3, and panel 5 shows the mask obtained by filling in the convex hull in panel 4.}
	\label{fig:clin_masks}
\end{figure*}

The clinician-provided segmentations were drawn as imprecise outlines on the ultrasound images (see Figure \ref{fig:clin_masks}), and therefore could not be used to generate masks in a simple and direct way. We therefore use OpenCV \cite{opencv} to convert these segmentations to masks. 

For each clinician-provided segmentation, we first compute its difference with the original unsegmented ultrasound image. Since some background noise iss retained in most images, we use an adaptive threshold to convert the difference image to a binary image $B(x,y)$ using the following formula:
    \begin{equation}
    B(x,y) = \begin{cases}
            0, & \text{if $S(x,y)>T(x,y)$}\\
            1, & \text{otherwise} 
        \end{cases}
    \end{equation}
where $T(x,y)$ is the mean in the $25\times40$ pixel neighbourhood around each pixel $(x,y)$ computed from the difference image $S(x,y)$.

We then use a border-following algorithm \cite{Suzuki1985} with Teh-Chin chain approximation \cite{Teh1989} to identify contours from the binary images. Contours with an area of less than 25 pixels are removed as noise. We compute and fill the convex hull of the remaining contours using the Sklansky algorithm \cite{Sklansky1982}. We use these as the ground truth masks for evaluating segmentation performance. 

Following this procedure, the masked images are visually inspected compared to the clinician-provided segmentations, and those masks which deviate significantly from the clinician's segmentations (e.g., because additional annotations are added to the segmented images, as seen in Figure \ref{fig:clin_masks}) are omitted from further analysis. In total, 53 images are removed (438 are used for evaluation), and the class distribution remains relatively even (range: 83-93 per class).

\section{Results}

\subsection{Synthetic Image Generation}
\begin{figure*}
	\centering
	\includegraphics[width=0.75\textwidth]{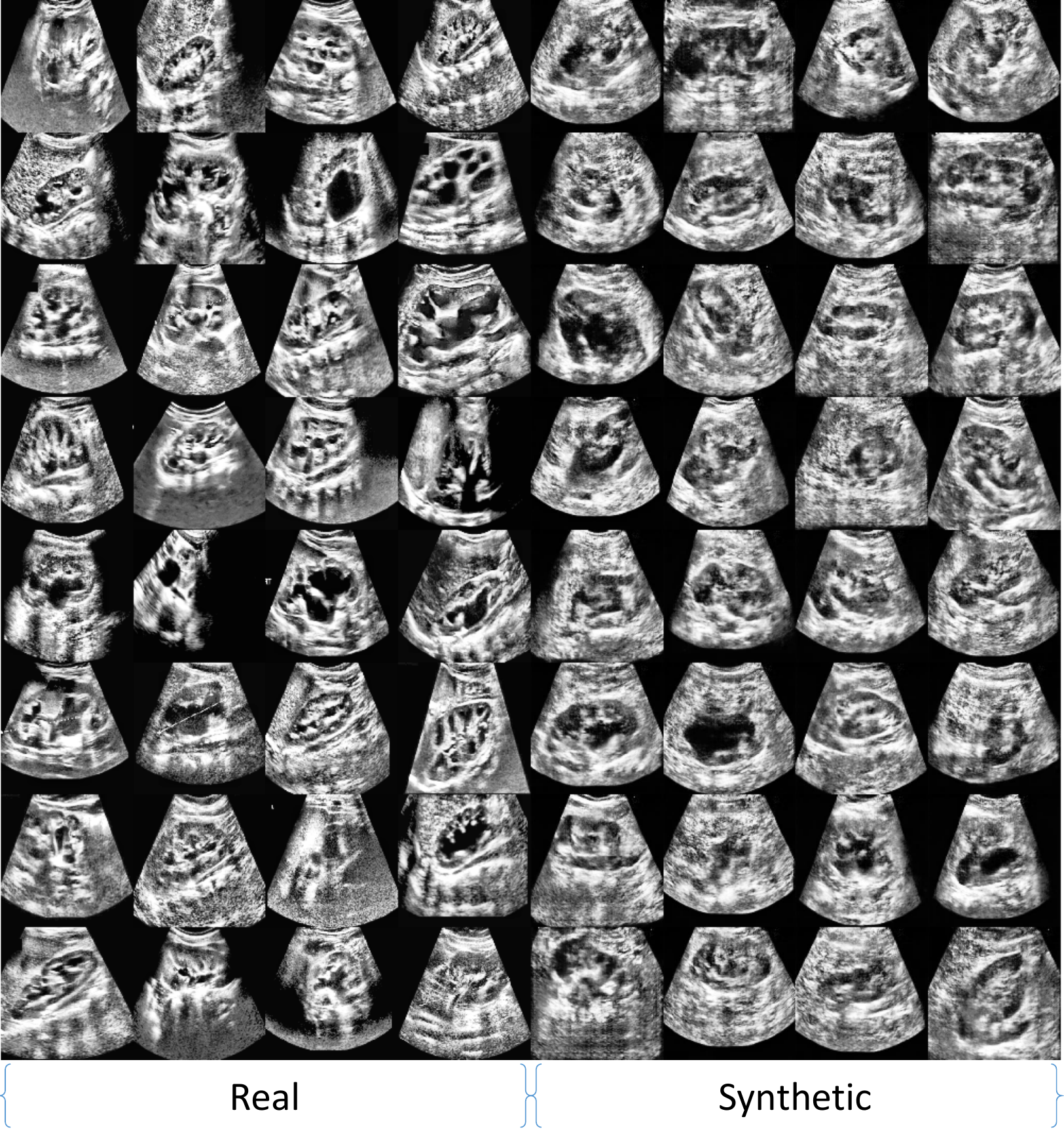}
	\caption{Real (first four columns) and generated (last four columns) kidney ultrasound images.}
	\label{fig:gen_kidney}
\end{figure*}

\begin{figure*}
	\centering
	\includegraphics[width=0.75\textwidth]{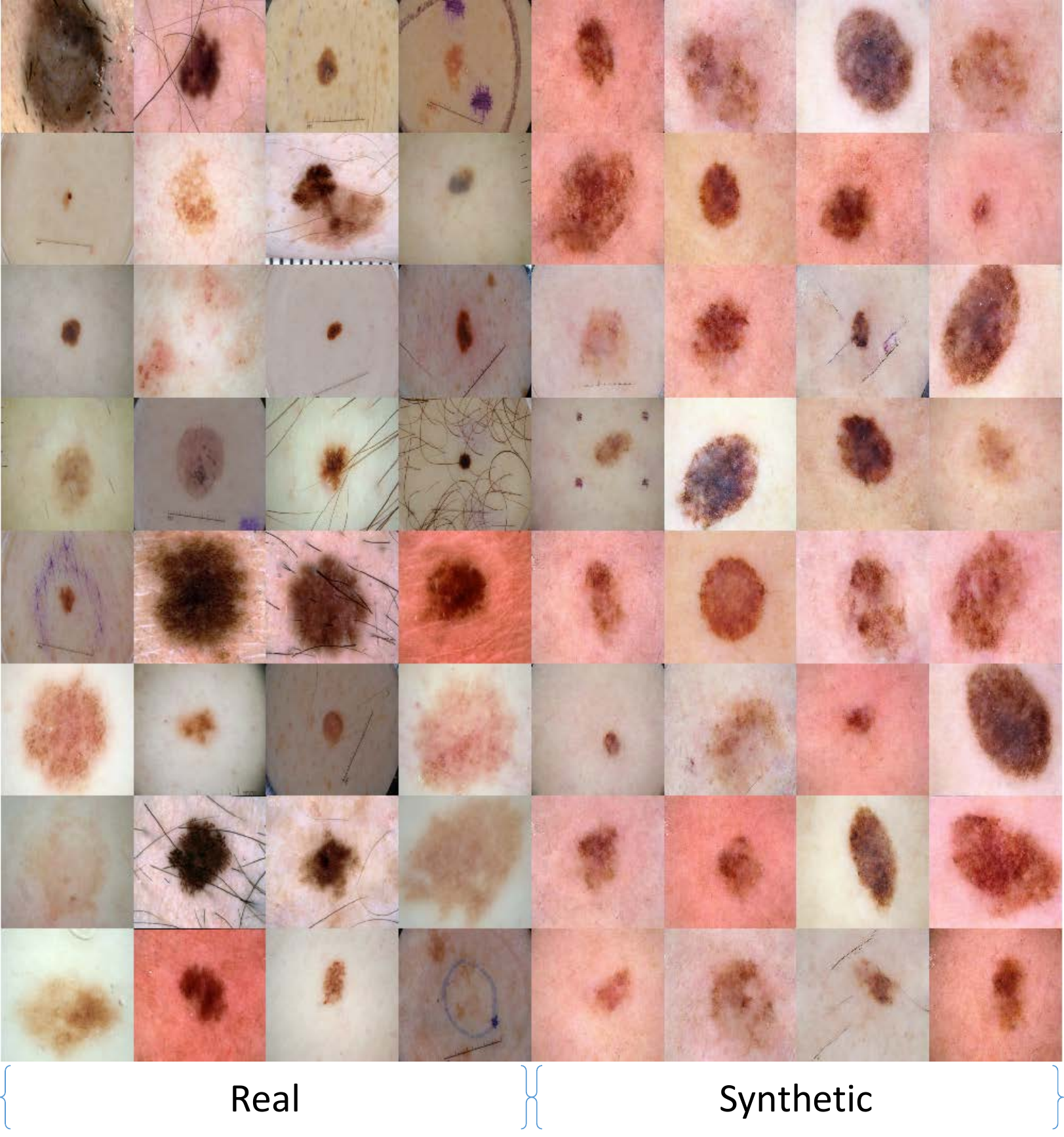}
	\caption{Real (first four columns) and generated (last four columns) dermoscopic images.}
	\label{fig:gen_skin}
\end{figure*}

To illustrate the similarity between real and synthetic images, we show a random sample of real and generated kidney ultrasound images in Figure \ref{fig:gen_kidney}, and a random sample of real and generated dermoscopic images in Figure \ref{fig:gen_skin}. Since our goal is to generate images that are similar enough for generalizeable training of a segmentation model, our approach does not produce state-of-the-art synthetic image generation. Instead, it produces images that have similar segmentation properties.

\subsection{Kidney Segmentation Performance}
\begin{figure*}
	\centering
	\includegraphics[width=0.9\textwidth]{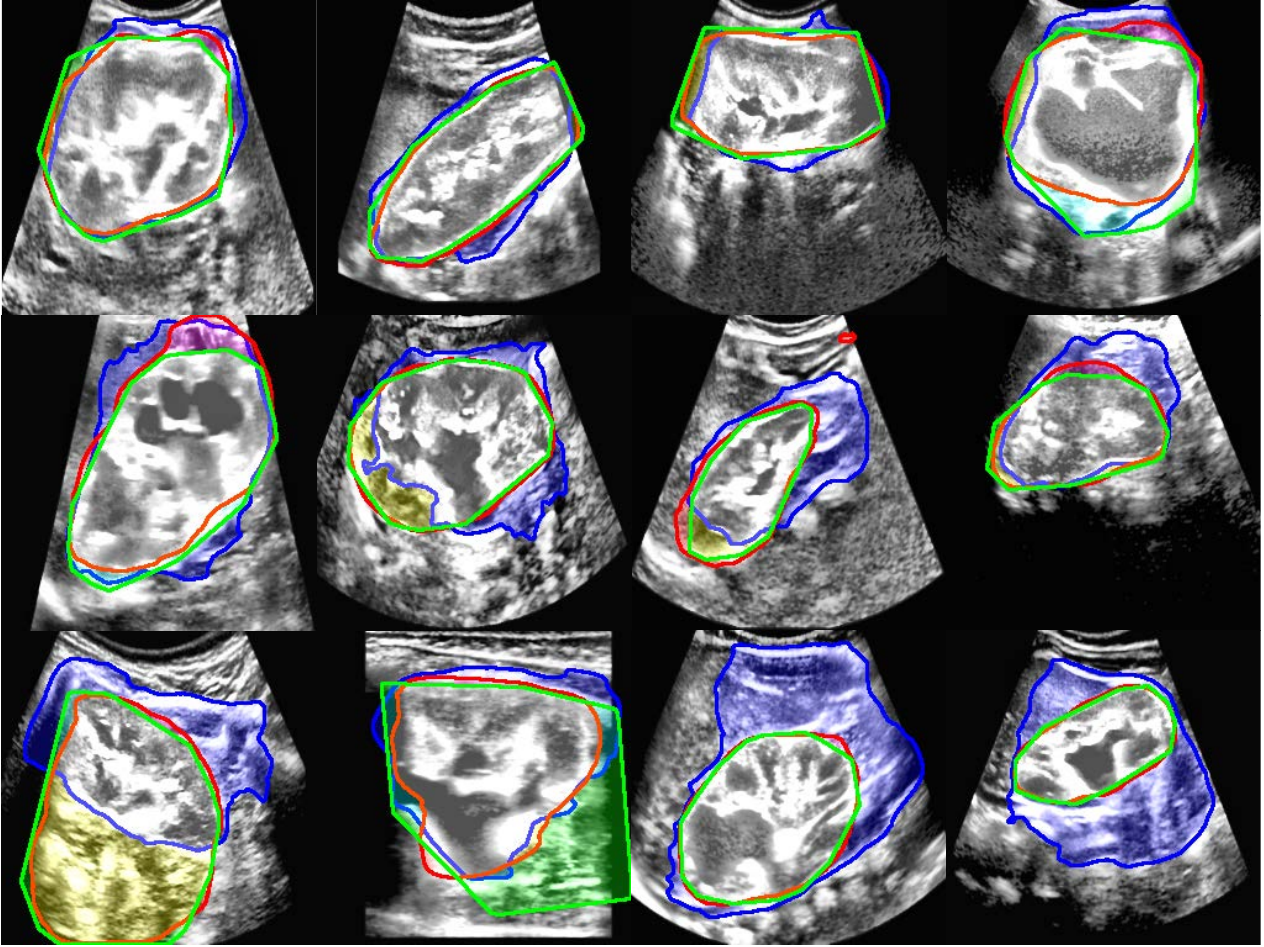}
	\caption{Kidney segmentation masks comparing our unsupervised method (blue) to a supervised U-Net (red) and the clinician-provided ground truth (green). Top row: images with high agreement between models. Middle row: images with moderate agreement between models. Bottom row: images with poor agreement between models (specific cases where the unsupervised approach fails).}
	\label{fig:qual_results}
\end{figure*}

In Figure \ref{fig:qual_results} we show the kidney segmentation masks learned through our fully unsupervised approach (with U-net as the segmentation model), compared with a purely supervised U-net and a purely unsupervised W-net. In Table \ref{tab:results} we show the corresponding segmentation performance metrics. For the semi-supervised extensions of our approach, we train a U-net using real and synthetic ultrasound images in a standard training protocol (U-net), and we also train a U-net using just the synthetic data followed by supervised fine-tuning with 45 of the real images with clinician segmentations, which were then removed from the evaluation set (U-net+). 

\begin{table*}
\begin{center}
	\caption{Performance metrics for ultrasound kidney segmentation.}
	\label{tab:results}
	\begin{tabular}{| c | c | c | c | c | c | c |}
	\hline
	{}  & Model & F1 & Specificity & Sensitivity & mIoU & pACC  \\ \hline
	\multirow{2}{*}{Unsup.}   
	& Ours (U-net)  & 0.81 (0.09) & 0.92 (0.05) & 0.87 (0.14) & 0.69 (0.12) & 0.90 (0.05) \\ 
	& W-net & 0.46 (0.10) & 0.20 (0.05) & 0.98 (0.02) & 0.41 (0.07) & 0.41 (0.07) \\ 
	 \hline
	\multirow{2}{*}{Semi-Sup.}
	& Ours (U-net)  & 0.87 (0.11) & 0.97 (0.04) & 0.86 (0.13) & 0.78 (0.13) & 0.93 (0.05) \\ 
	& Ours (U-net+) & 0.88 (0.08) & 0.97 (0.03) & 0.88 (0.09) & 0.80 (0.11) & 0.94 (0.04) \\ 
	\hline
	\multirow{1}{*}{Sup.}     
	& U-net & 0.91 (0.09) & 0.97 (0.04) & 0.90 (0.10) & 0.84 (0.10) & 0.95 (0.03) \\ 
	\hline
	\end{tabular}
\end{center}
\end{table*}

\subsection{Skin Lesion Segmentation Performance}
\begin{figure*}
	\centering
	\includegraphics[width=0.9\textwidth]{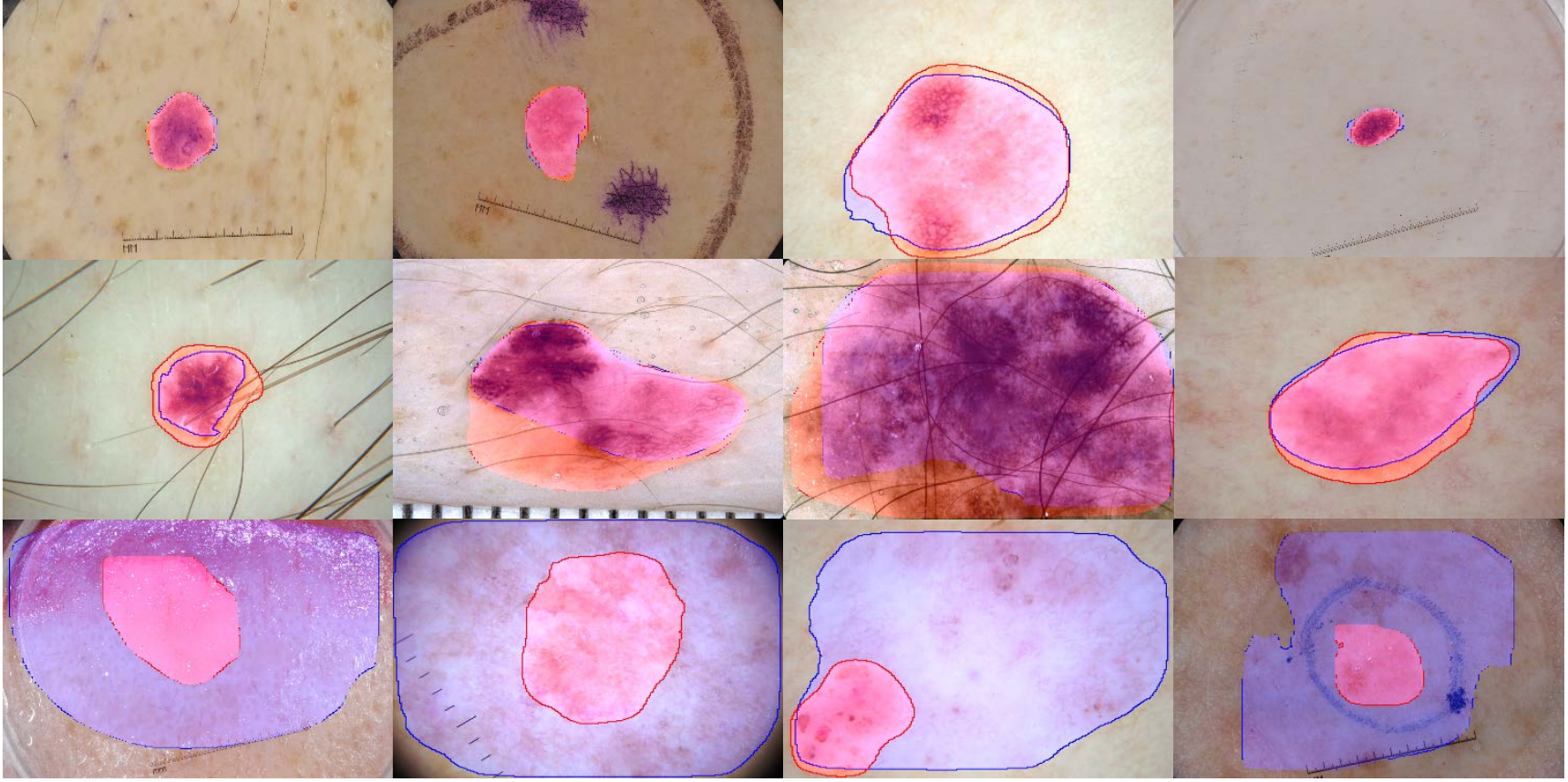}
	\caption{Skin lesion segmentation masks comparing our unsupervised method (blue) to a supervised Mask-RCNN (red). Top row: images with high agreement between models. Middle row: images with moderate agreement between models. Bottom row: images with poor agreement between models (specific cases where the unsupervised approach fails). Ground truth segmentation not available for ISIC 2018 test images.}
	\label{fig:qual_results_skin}
\end{figure*}

Performance metrics on the ISIC 2018 dataset using our unsupervisd approach are shown in Table \ref{tab:results_skin} along with results obtained by the competition winner and current top submission. Here we use the metrics given by the online submission system, which includes a thresholded mIoU (th-mIoU). This metric sets all per-image IoU scores that are less than 0.65 to 0 before computing the mean IoU. Semi-supervised results are not available because the ISIC 2018 test submission page has been removed whie preparing this manuscript, and the test set is not currently available. Examples of the output masks on validation images are shown in Figure \ref{fig:qual_results_skin}.

\begin{table*}
\begin{center}
	\caption{Performance metrics for ISIC 2018 skin lesion boundary segmentation.}
	\label{tab:results_skin}
	\begin{tabular}{| c | c | c | c | c | c | c | c |}
	\hline
	{}  & Model & F1 & Specificity & Sensitivity & mIoU & th-mIoU & pACC  \\ \hline
	\multirow{2}{*}{Unsup.}   
	& Ours (Mask-RCNN) & 0.830 & 0.947 & 0.835 & 0.753 & 0.683 & 0.904 \\ 
	& Ali et al. 2019 \cite{Ali2019} & 0.543 & n/a & n/a & 0.440 & n/a & n/a \\
	\hline
	\multirow{3}{*}{Sup.}     
	& Mask-RCNN & 0.882 & 0.950 & 0.922 & 0.811 & 0.763 & 0.936 \\ 
	& Winner & 0.898 & 0.963 & 0.906 & 0.838 & 0.802 & 0.942 \\ 
	& Current Top & 0.915 & 0.941 & 0.956 & 0.852 & 0.836 & 0.954 \\ 
	\hline
	\end{tabular}
\end{center}
\end{table*}

\section{Discussion}
We present an unsupervised approach to semantic medical image segmentation that takes advantage of recent advances in image synthesis and generative modelling by making assumptions about the common geometry inherent to an object of interest. This method performs better than some previous unsupervised methods that fit the problem definition (e.g., W-net), or for which results are available (e.g., the CNN-based approach in \cite{Ali2019}). For example, W-net performs poorly on the kidney segmentation task because it only identifies the ultrasound cone itself, rather than the kidney. We also show that our approach performs nearly as well as supervised methods for most images. Importantly, we show that with just a few training examples for supervised fine-tuning (here, only 10\% of the data used for the supervised models), we approach the segmentation performance of purely supervised models. 

Our method tends towards identifying larger ROIs that contain the desired ROI, which results in high specificity (0.92 and 0.947 for the kidney dataset and ISIC 2018 respectively) and only moderate sensitivity (0.87 and 0.835). For both datasets, the model fails for a small subset of the images. In the case of the kidney dataset, we find no clear pattern to explain the failed images. However, in the case of ISIC 2018 images, the unsupervised model does poorly with images that contain a lens or film placed on top of the skin lesion (in these cases, the model incorrectly segments the lens instead of the skin lesion underneath). 

Interestingly, even though we construct edge diagrams based on smooth and convex shapes for image synthesis, the resulting segmentation models are able to fit non-smooth and non-convex boundaries. It is possible that alternative methods for generating edge diagrams with greater complexity may lead to a more flexible model that can adapt to more complex geometries. We are currently exploring the utility of this method in segmenting organs with more complex geometries, segmenting multiple objects per image, and performing 3D segmentation. We are also currently exploring adaptations of our approach that make it more end-to-end, e.g., by using multiple GANs.



{\small
\bibliographystyle{ieee_fullname}
\bibliography{References}
}

\end{document}